\documentclass[journal]{IEEEtran}
\usepackage{amsmath,amssymb,amsfonts,tabularx}
\usepackage[hidelinks]{hyperref}
\usepackage{cite}
\usepackage{url}
\usepackage{paralist}
\usepackage{flushend}
\usepackage{graphicx}
\usepackage{makecell}
\usepackage{xcolor}
\usepackage{array,multirow}
\usepackage{xspace}
\graphicspath{{figures/}}
\usepackage{adjustbox, makecell, colortbl, xcolor}
\usepackage{amssymb}
\usepackage{algorithm}
\usepackage{algpseudocode}
\usepackage{tabularx}
\usepackage{booktabs}
\usepackage{array}
\usepackage{tcolorbox}
\usepackage{adjustbox, colortbl}
\usepackage{algpseudocode}
\usepackage{booktabs}
\usepackage{placeins}
\usepackage{mdframed}

\newenvironment{remarkbox}%
{%
  \begin{mdframed}[%
    backgroundcolor=gray!10,
    linecolor=white,
    linewidth=0pt,
    skipabove=\topskip,
    skipbelow=\topskip
  ]
  \bfseries TAKEAWAY.~\normalfont
}%
{%
  \end{mdframed}
}

\newenvironment{takeawaybox}%
{%
  \begin{mdframed}[%
    backgroundcolor=gray!10,  
    linecolor=white,          
    linewidth=0pt,            
    skipabove=\topskip,       
    skipbelow=\topskip,       
    innerleftmargin=6pt,      
    innerrightmargin=6pt,     
    innertopmargin=6pt,       
    innerbottommargin=6pt     
  ]
  \textbf{LIMITATIONS.}\ \normalfont  
}%
{%
  \end{mdframed}
}

\definecolor{lightgray}{gray}{0.9}
\definecolor{verylightblue}{RGB}{230, 245, 255}

\newcommand{\cmark}{\checkmark} 
\newcommand{\xmark}{\text{\sffamily X}} 

\begin{document}

\title{Vulnerability Disclosure through Adaptive Black-Box Adversarial Attacks on NIDS}
\author{
    \IEEEauthorblockN{
        Sabrine Ennaji\IEEEauthorrefmark{1},
        Elhadj Benkhelifa\IEEEauthorrefmark{2},
        Luigi V. Mancini\IEEEauthorrefmark{1}\\
    }
    \IEEEauthorblockA{
        \IEEEauthorrefmark{1}Sapienza University of Rome, 
        \{ennaji, mancini\}@di.uniroma1.it\\
    }
    \IEEEauthorblockA{
        \IEEEauthorrefmark{2}University of Staffordshire, 
        e.benkhelifa@staff.staffs.ac.uk\\
    }
}

\maketitle

\begin{abstract}

Adversarial attacks, wherein slight inputs are carefully crafted to mislead intelligent models, have attracted increasing attention. However, a critical gap persists between theoretical advancements and practical application, particularly in structured data like network traffic, where interdependent features complicate effective adversarial manipulations. Moreover, ambiguity in current approaches restricts reproducibility and limits progress in this field. Hence, existing defenses often fail to handle evolving adversarial attacks. This paper proposes a novel approach for black-box adversarial attacks, that addresses these limitations. Unlike prior work, which often assumes system access or relies on repeated probing, our method strictly respect black-box constraints, reducing interaction to avoid detection and better reflect real-world scenarios. We present an adaptive feature selection strategy using change-point detection and causality analysis to identify and target sensitive features to perturbations. This lightweight design ensures low computational cost and high deployability. Our comprehensive experiments show the attack's effectiveness in evading detection with minimal interaction, enhancing its adaptability and applicability in real-world scenarios. By advancing the understanding of adversarial attacks in network traffic, this work lays a foundation for developing robust defenses.

\end{abstract}

\IEEEpeerreviewmaketitle

\section{Introduction}
\IEEEPARstart{I}n today’s interconnected world, almost every aspect of our personal and professional lives is dependent on the digital infrastructure. 
Moreover, the growing volumes of sensitive data transmitted over networks present significant threats to privacy, security, and even national stability. Intrusion detection systems (IDS) play a vital role in safeguarding networks by continuously monitoring the network traffic for suspicious activities \cite{ozkan2021comprehensive}. However, the rising complexity of cyber threats exposes the limitations of traditional IDS in detecting sophisticated attacks due to their reliance on predefined signatures. To fill this void, integrating machine learning (ML) with IDS has revolutionized network security, offering proactive detection by analyzing large traffic datasets, learning patterns, and identifying potential breaches \cite{chou2021survey}. 

Critically, the increasing adoption of ML-based IDS has exposed them to adversarial attacks, where slight perturbations could easily trick
ML models into making wrong predictions \cite{shu2020generative}. This threat has been initially developed in the image classification context (i.e., unstructured data) without considering specific feature importance \cite{goodfellow2014explaining, zhang2019adversarial}. In contrast to ML-based IDS that rely on structured network traffic, each perturbation must adhere to the inherent constraints of the features since they hold significant semantic values and are intricately interconnected \cite{rigaki2018bringing}. Any random perturbation that does not respect these constraints is likely to be detected as anomalous or unrealistic, decreasing the effectiveness of the attack \cite{ennaji2024adversarial}. Consequently, developing successful adversarial attacks within network traffic necessitates a deep understanding of network protocols, feature interdependencies, and the specific vulnerabilities within the model's architecture.

Despite extensive research in adversarial machine learning, available countermeasures show limited effectiveness in practical applications and poor generalizability \cite{apruzzese2022modeling}. This arises from the focus of most existing studies on white-box attacks, which are developed under controlled conditions, without considering the dynamic, complex, and noisy environments in which IDS operate \cite{mohammadian2023gradient, alhajjar2021adversarial}. Additionally, the limited literature for adversarial attacks in black-box settings, which better reflect real-world scenarios, lacks clarity on how attackers gather information about the target IDS and what techniques are used to extract responsible features for the model's decision-making \cite{sheatsley2022adversarial, vitorino2022adaptative}. This ambiguity and disconnect between theoretical models and practical implementation restricts the understanding of researchers and industry practitioners, adding to the complexity of designing robust defenses in this evolving field. 

Microsoft researchers have interviewed 28 organizations across various domains (e.g., cybersecurity, healthcare, government, banking, agriculture) and all of them confirmed this inadequacy and discussed challenges in applying academic research findings to real-world scenarios \cite{kumar2020adversarial}. They expressed serious concerns about expanding research beyond traditional threats (e.g., phishing and malware) to cover real-world adversarial manipulation challenges. 

Building on these critical gaps, this research work advances the state-of-the-art in adversarial attacks by proposing a novel adaptive technique that effectively bypasses Network-IDS (NIDS). The main contributions of this study are as follows.  



\begin{enumerate}
\item \textbf{Real-world vulnerability assessment:} Our proposed strategy strictly adheres to black-box constraints, where attackers cannot directly access the target model or prior knowledge of its internal workings. It relies only on indirect observations with minimal interaction, reducing detection risks and simulating realistic real-world scenarios. In contrast to existing black-box adversarial attacks that often require multiple queries to the IDS to gather information, increasing the likelihood of being detected. Moreover, we consider network complexity, feature interdependencies, and dynamic environments, enhancing the approach's feasibility in real-world deployments.
\item \textbf{Blind selection of sensitive features:} We propose a novel strategy for blind selection of sensitive features using change-point detection and causality analysis. This helps identify the most susceptible features to perturbations without direct system access, which is critical for successful adversarial attacks in black-box settings.
\item \textbf{Lightweight and highly deployable design:} Our approach is designed to be lightweight, minimizing computational requirements and facilitating integration into existing IDS frameworks. 
\item \textbf{Foundation for developing robust defenses:} We raise awareness about the silent probing threat and why available adversarial countermeasures fail in practice. By enhancing the understanding of how adversaries exploit vulnerabilities, this study provides a foundation for the research community to develop stronger and more resilient defenses, addressing industry needs.
\end{enumerate}

This paper is structured as follows: Section \textbf{2} provides an overview of the background, introducing key concepts related to adversarial attacks and defining the assumptions of black-box settings.
Section \textbf{3} critically reviews related work, highlighting the limitations of existing adversarial attack strategies and explaining why existing defenses fail in practice. 
Section \textbf{4} presents the dataset, target model and our proposed methodology for vulnerability assessment in black-box adversarial attacks, emphasizing an adaptive feature selection strategy using change-point detection and causality analysis to identify sensitive features. 
Section \textbf{5} provides a comprehensive analysis of the findings in comparison with existing approaches in the literature. 
Finally, Section \textbf{6} concludes the paper by summarizing our key contributions and discussing future directions.


\section{Fundamentals and Realistic Constraints for Adversarial Attacks} 
\label{sec:prerequisites}
This section lays the groundwork for our discussion by providing a simplified taxonomy of adversarial machine learning. It also presents the established methods for vulnerability assessment in adversarial black-box scenarios, exploring the challenges and key factors for realistic adversarial attacks in NIDS.  For a more comprehensive exploration, we encourage readers to consult the references \cite{ennaji2024adversarial, apruzzese2022modeling, pinto2023survey, alotaibi2023adversarial, martins2020adversarial}.

 \begin{figure*}
  \centering
  \includegraphics[trim= 4.5cm 21cm 3cm 0cm, clip, width=1\textwidth]{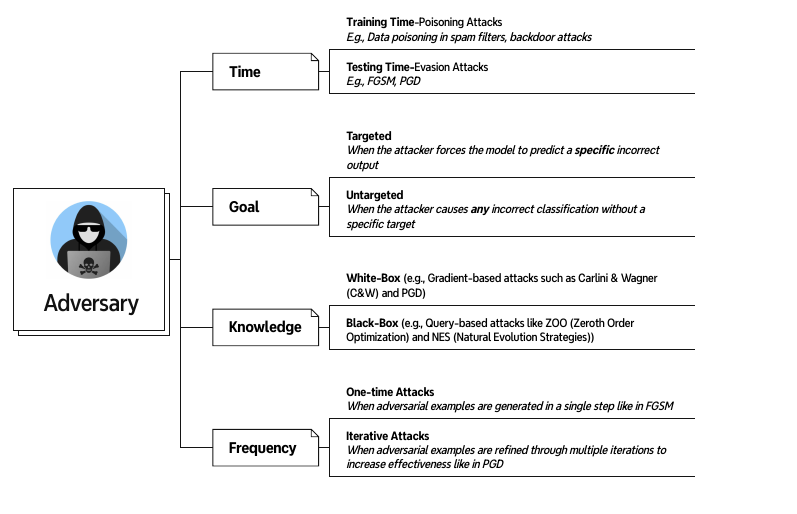}
  \caption{Adversarial Machine Learning Taxonomy}
\label{fig:taxonomy}
\end{figure*}

\subsection{Adversarial Machine Learning Taxonomy}
\textbf{Adversarial Threat.} 
Generally, an adversarial attack is a strategy that aims to fool a ML model into making wrong predictions by slightly modifying its input data \cite{michel2022survey}.

Mathematically, this threat aims to find a small perturbation \( \delta \) that, when injected to a legitimate input \( x \), results in an adversarial example \( x' = x + \delta \). The latter is generated to maximize the model's error, leading it to misclassify \( x' \) while appearing almost similar to the original input \( x \).

For a classifier \( f \) and a target class \( y \), the aim is to solve the following optimization problem:

\[
\text{max}_{\delta} \, \mathcal{L}(f(x + \delta), y) \quad \text{subject to} \quad \| \delta \| \leq \epsilon
\]

where:\\
 \( \mathcal{L} \) is the model loss function of the model\\
 \( \| \delta \| \leq \epsilon \) ensures the perturbation \( \delta \) remains minor

Therefore, this causes the model's output \( f(x') \) to diverge from \( f(x) \), mismatching \( x' \) with its original label \( y \) and making it adversarial.

There are four crucial elements that present and contribute to the effectiveness of adversarial attacks, as summarized in Figure \ref{fig:taxonomy}.

\begin{enumerate}
\item \textbf{Adversary timing:} Adversarial manipulations can occur during training or testing. Training-time attacks, known as poisoning attacks, degrade the learning process of the model by injecting malicious data. In contrast, testing-time attacks, known as evasion attacks, target a fully trained model by manipulating specific inputs to force incorrect predictions.
\item \textbf{Adversary goal:} Adversarial attacks can be targeted or non-targeted. While non-targeted attacks only seek to force any inaccurate prediction, targeted attacks aim to induce a specific incorrect prediction. These two types of classification converge in binary classification.
\item \textbf{Adversary knowledge:} This refers to the level of information an attacker has about the target system, typically categorized into white-box and black-box attacks.
\begin{itemize}
    \item \textbf{\textit{white-box attacks:}} The attacker has full access to the internal workings of the model, including the architecture, training data, parameters and gradients (e.g., FGSM \cite{papernot2016limitations})
    \item \textbf{\textit{black-box attacks:}} Without knowing how the model operates inside, the attacker can only access its outputs, such as predictions or confidence scores (e.g., ZOO \cite{chen2017zoo}, HopSkipJump \cite{chen2020hopskipjumpattack})
\end{itemize}
\textbf{Adversary frequency:} Adversarial attacks can be one-time or iterative attacks. One-time attacks generate adversarial examples in a single step (e.g, FGSM \cite{papernot2016limitations}), while iterative attacks refine these examples through multiple iterations (e.g., PGD \cite{madry2017towards}). 
\end{enumerate}

\textbf{Adversarial Defenses.} Significant advancements have been made in adversarial defenses, but none have proven to be completely effective \cite{ennaji2024adversarial}. Current defenses can be broadly categorized into two main approaches: Adversarial training and model architecture reinforcement. 
\begin{itemize}
\item \textbf{Adversarial Training:} It improves the robustness of the model by incorporating adversarial examples into its training process. Mathematically, it aims to minimize the loss function \( \mathcal{L} \) over both clean and adversarially perturbed data. To counter adversarial attacks, which specifically maximizes the model’s loss within a limited perturbation rage, adversarial training minimizes the expected loss under the highest-impact perturbation. Given an input \( x \) with label \( y \), this purpose can be formulated as:

\[
\min_{\theta} \, \mathbb{E}_{(x, y) \sim D} \left[ \max_{\delta : \|\delta\| \leq \epsilon} \, \mathcal{L}(f_{\theta}(x + \delta), y) \right]
\]

The inner maximization finds the highest-impact perturbation \( \delta \) within the allowed range \( \epsilon \), and the outer minimization adjusts model parameters \( \theta \) to minimize this loss. 

\item \textbf{Model Architecture Reinforcement:} This refers to improving the structure and inherent robustness of the model to make it more resilient to adversarial attacks, reducing reliance on adversarial training and helping the model detect outliers and perturbations. Different techniques can be included such as,
ensemble learning \cite{pang2019improving}, detection and rejection mechanisms, which identify and filter out adversarial examples \cite{metzen2017detecting}, input transformation strategy \cite{ali2022evaluating}, gradient masking \cite{chen2022adversarial} and manifold projection \cite{ma2018characterizing}.
\end{itemize}

\begin{table*}[h!]
\centering
\renewcommand{\arraystretch}{1.5}
\rowcolors{2}{gray!15}{white}
\caption{Existing Vulnerability Assessment for Adversarial Attacks}
\label{tab:vulnerability_methods}
\begin{tabular}{l p{10.5cm} c}
    \rowcolor{gray!40}
    \textbf{Vulnerability Assessment} & \textbf{Limitations} & \textbf{Real-World Feasibility} \\
    \hline
    Query-based & High query costs, Easily detected, Time-consuming & Low \\
    Decision Boundary-based & Computationally expensive, Requires precise model boundary approximations & Medium \\
    Randomized \& Gradient-Free & Requires extensive sampling, High computational overhead & High \\
    Transferability & Surrogate model mismatch, High dependency on data similarity, Less effective against robust models & Medium \\
\end{tabular}
\end{table*}

\subsection{Vulnerability Assessment for Black-Box Adversarial Attacks}
In the context of network traffic, identifying the most sensitive features is crucial, as these features are often the primary focus in adversarial attacks. Additionally, their manipulation heavily relies on preserving realistic constraints and respecting the inherent correlations between them. Ignoring these interdependencies frequently results in attacks that are either impractical or unrealistic. Existing vulnerability assessment methods frequently overlook these critical aspects, leading to evaluations that lack real-world relevance. This gap limits the understanding of adversarial attacks and the development of robust defenses that generalize effectively. These assessments can be categorized into four groups:
\begin{enumerate}
\item \textbf{Query-based method:} This involves directly querying the target model to obtain feedback, then creating adversarial examples based on the observed responses \cite{bai2023query}.
\item \textbf{Decision boundary-based method:} By examining the model's outputs for carefully selected inputs, these methods employ optimisation techniques to estimate the model's decision boundaries, aimimg to generate adversarial examples that undetectedly cross them \cite{finlay2019logbarrier}.
\item \textbf{Randomized and gradient-free method:} It is appropriate for models with inaccessible gradients since they rely on random perturbations or optimization strategies that do not require gradient information \cite{xiang2023improving}.
\item \textbf{Transferability:} It exploits the transferability of adversarial examples, demonstrating that perturbations created for a specific target model can also mislead other models with different architectures \cite{wang2023towards}.
\end{enumerate}

Table \ref{tab:vulnerability_methods} depicts the limitations and real-world feasibility of current vulnerability assessment methods, which are addressed by our proposed framework and will be compared to it in Section \ref{sec:discussion}.

\subsection{Challenges and Key Factors for Realistic Adversarial Attacks in Network Intrusion Detection Systems}
\begin{figure*}
  \centering
  \includegraphics[trim= 0cm 0cm 0cm 0cm, clip, width=0.77\textwidth]{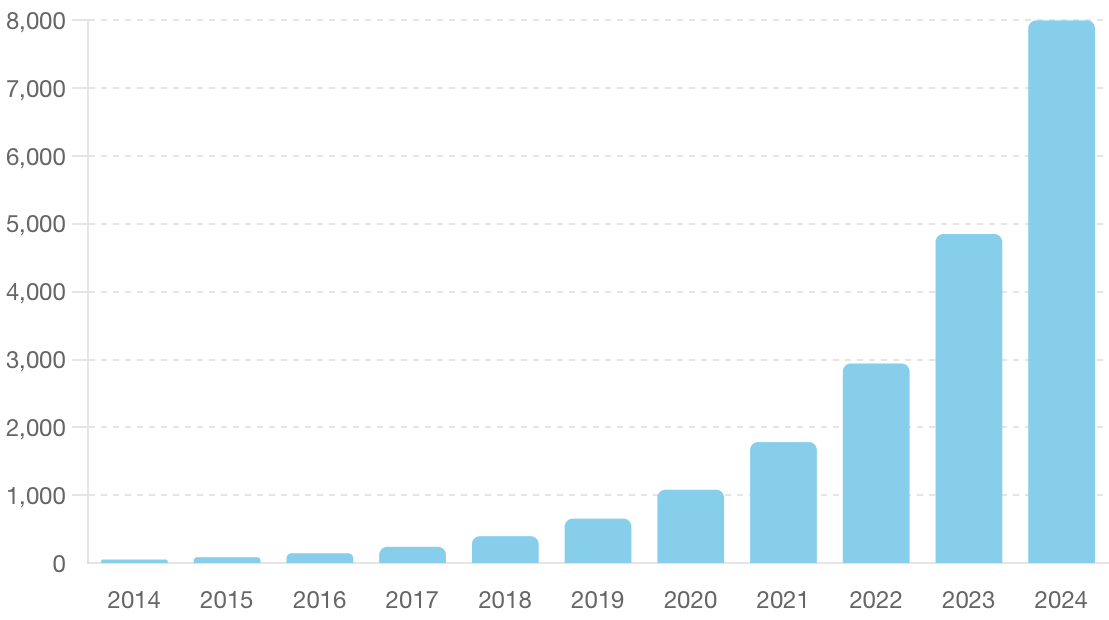}
  \caption{Adversarial Machine Learning Growth Over Years \cite{carlini2019complete}}
\label{fig:research}
\end{figure*}

Despite growing research (Figure \ref{fig:research}), many adversarial attacks in NIDS fail to consider real-world constraints like feature interdependencies in network traffic, as demonstrated in section \ref{sec:review}. Furthermore, assumptions about attackers having full access to training data or building substitute models in black-box settings often lack practical feasibility \cite{apruzzese2022modeling}. Many black-box attacks assume unrestricted oracle access to the target NIDS. This disconnect between theory and practice limits their real-world applicability.

To ensure that assessments of adversarial attacks consider realistic restrictions and that countermeasures are effective in real-world scenarios, it is essential to address these factors. 
\begin{itemize}
\item \textbf{Data accessibility:} The relationship between an attacker's level of access to training data (no access, partial access, or full access) and the resulting impact on the robustness of NIDS.
\item \textbf{Feature Constraints:} The extent to which adversarial perturbations maintain the real-world correlations between features in network traffic.
\item \textbf{Response Interaction:} The reliance on model outputs (e.g., predictions or confidence scores), and the limitations on the amount of queries an attacker can make in practical deployment scenarios.
\item \textbf{Perturbation Domain:} Whether adversarial perturbations are generated directly to raw network traffic or to the higher-level feature representations used by the NIDS.
\item \textbf{Model Insight:} The level of information the attacker has about the internal workings of the target NIDS, and how different levels of model understanding influence attack feasibility and defense mechanisms.
\end{itemize}
Advancing this field requires the creation of threat models that align with real-world constraints and the evaluation of adversarial robustness within such practical scenarios.

\section{Literature Review}
\label{sec:review}
This study is motivated by three main factors: the rising adversarial threat to intelligent systems, the overreliance on white-box attack methodologies, and the limited development of robust black-box strategies that respect practical conditions. This section critically reviews existing work in both areas; white-box and black-box settings, which is comprehensively summarized in Table \ref{tab:gen_adv}. Additionally, it identifies the open research questions \textit{(limitations)} addressed in this paper.

\subsection{White-Box Settings}
The majority of research on adversarial attacks against ML-based NIDS has focused on white-box settings, in which attackers have complete knowledge of the target model's architecture, parameters, and training data. For instance in \cite{mohammadian2023gradient}, the authors employs the Jacobian Saliency Map Attack (JSMA) to selectively perturb features of a Multi-Layer Perceptron (MLP) based-NIDS. It is an adversarial attack initially proposed by Papernot et al.  \cite{papernot2016limitations}, but in the image classification context. JSMA identifies key features based on their saliency values, which indicate how sensitive they are  to alterations in the model’s output. By calculating derivatives, this method helps selecting which features to minimally perturb to mislead the model. The proposed findings show effectiveness across different datasets. However, in network intrusion detection, it is mandatory to respect domain-specific constraints (e.g., timing correlations, packet validity, protocol adherence, etc.) and the paper lacks clarity on how authors addressed this requirement to maintain realistic and valid traffic characteristics. 

In contrast to Sheatsley et el. \cite{sheatsley2022adversarial}, they explicitly integrated network constraints into the adversarial crafting process of the traditional JSMA against DNN-based NIDS. This adaptive version (AJSMA) integrates protocol-specific rules to ensure that adversarial manipulations considers network semantics, such as maintaining the integrity of TCP/IP protocol features. Moreover, it provides dynamic perturbation directions to optimize the selection of significant features for perturbation. Therefore, it maximizes the attack’s effectiveness while preserving the validity of network traffic. The authors also improve the generalization of their approach over varying attack models by presenting a Histogram Sketch Generation (HSG) strategy that constructs universal adversarial perturbations based on the frequency and impact of feature modifications for different inputs. However, the methodology shows ambiguities on how network constraints are enforced and adversarial examples are validated.

Another extension of JSMA by Anthi et al. \cite{anthi2021adversarial}, has been explored in targeting Random Forest (RF) and J48-based NIDS within an Industrial Control Systems (ICS) environment. Since JSMA depends on gradient information to identify and modify sensitive features, it cannot be directly applied to non-differentiable classifiers like RF and J48. Hence, a pre-trained DNN has been used as a surrogate model to transfer adversarial samples to the target models. While this transferability concept effectively shows the susceptibility of gradient-independent models to JSMA, the study does not address feature consistency and semantic validity. 

Similarly, in a recent study \cite{roshan2024untargeted}, the authors conducted a real-time adversarial attack simulation with different traditional white-box attacks like FGSM, JSMA, PGD, and C\&W, injecting perturbed network packets back into the traffic flow to assess the resilience of DNN-based NIDS models. However, they do not sufficiently explain and address the challenge of preserving the realistic and functional nature of network traffic features. This raises questions about the validity of the achieved results. 

Roshan et al. have proposed in another study \cite{roshan2024boosting} a two-phase adversarial defense to improve the robustness of DNN-based NIDS against white-box adversarial attacks, specifically the (C\&W) attack that relies on an optimization-based approach to create imperceptible alterations while balancing the misclassification rate. Their defense integrates Gaussian Data Augmentation (GDA) during training to inject noise and enhance the model's resilience and Feature Squeezing (FS) during testing to minimize the input’s resolution and mitigate adversarial effects. While this defense shows effectiveness against C\&W, the authors do not focus on explaining the adopted feature selection method, which is crucial for validating whether their attack and countermeasure procedures remain logical and align with dynamic real-world conditions. Moreover, further experimentation is needed to prove the generalizability of the proposed defense to other adversarial attacks.

In the same way, Pawlicki et al. \cite{pawlicki2020defending} propose a defense strategy against four adversarial attacks (FGSM, BIM, C\&W, and PGD) targeting a DNN-based NIDS. They used neural activations at the test time to detect adversarial examples. Fundamentally, the activations of each layer of the neural network are recorded for both normal and adversarial inputs, creating an activation dataset. Based on the latter, the authors trained an additional ANN-based detector to differentiate between adversarial and non-adversarial samples. Like most existing proposals in the literature, this study does not explicitly explain how feature interdependencies have been preserved, nor does it address semantic validity in the generated attacks. This raises concerns about the feasibility of the detected samples in real-world scenarios. Furthermore, the reliance on neural activations could be susceptible to adversarial attacks that manipulate these patterns. This method could also introduce trade-offs regarding processing time and detection latency and potentially increase false positives. 

Additionally, Costa et al. \cite{costa2024argan} have recently used FGSM attack to mislead KitNET, a lightweight anomaly detection system that leverages an ensemble of autoencoders. Each autoencoder focuses on recreating specific network traffic features, with a larger output autoencoder that can handle more complex patterns. While FGSM effectively fools KitNET by injecting subtle perturbations into the input that lower the reconstruction error, it perturbs all features equally based on the gradient direction without explicitly selecting sensitive features. To defend against this attack, the authors propose ARGAN-IDS, a GAN-based defense that reconstructs adversarial inputs into more "benign-looking" forms. However, given that the evaluated adversarial attack does not consider the interdependency and complexity of network features, challenges remain to ensure that ARGAN-IDS accurately authenticates the reconstructed samples within practical scenarios.

\begin{takeawaybox} 
While these assumptions demonstrate the vulnerability of ML models, they rarely hold in NIDS real-world settings, where system details are known only to internal system administrators \cite{apruzzese2022modeling}. Even the companies purchasing and deploying these systems often lack full knowledge of their inner workings. Thus, it is highly unlikely that attackers will have the extensive access required for white-box attacks in practice. Moreover, they frequently ignore feature constraints, such as maintaining network protocol rules, feature interdependencies, and semantic consistency \cite{ennaji2024adversarial}. This limits the practical applicability of these attacks and restricts the development of reliable defenses.
\end{takeawaybox}

\subsection{Black-Box Settings}
While a few existing proposals consider black-box adversarial attacks to align more closely with practical conditions, only a few truly adhere to a genuine black-box setting. 
In \cite{zhang2022adversarial}, a variety of black-box attacks (e.g., Natural Evolution Strategies (NES), Boundary Attack, HopSkipJumpAttack, Pointwise Attack, and OPT-Attack) have been evaluated against different NIDS architectures; C-LSTM, an ensemble model combining CNN and LSTM, and individual DNN and CNN-based NIDS. Their impact has been assessed by measuring their success rate, query efficiency, and transferability on both individual and ensemble models. As a proposed defense against these attacks, the authors relied on three strategies: Voting Ensembling technique to reinforce the model's decision-making; Ensemble Adversarial Training (EAT), where models are trained using adversarial examples generated from several attacks; and Adversarial Query Detection, which examines repetitive and similar traffic patterns to detect adversarial probing. Despite the promising results, the authors focus more on the effectiveness of the attacks and ignore several key areas, such as the methods attackers would use to identify the sensitive features to perturb and how they gathered information about the target model. 

On the other hand, Peng et al. \cite{peng2019adversarial} address in their study some feature constraints by preserving feature ranges. They present an optimization-based attack against a DNN-based NIDS that relies on iteratively perturbing continuous and discrete features using a gradient-free approach and random walk strategy to misclassify samples while maintaining similarity to original DoS traffic. However, this paper fails to provide a comprehensive explanation of how the feature dependencies and protocol-specific rules have been considered. Additionally, the feature selection process and the technique for gathering information about the target model are not clearly detailed. This limits the practical assessment of the attack.

Furthermore, in \cite{wu2019evading} a black-box attack framework has been proposed based on deep reinforcement learning (DRL) to bypass botnet detection systems targeting a CNN and a decision tree (DT)-based models. It uses a Deep Q-Network (DQN) agent to perturb significant network flow features (e.g., packet timestamps and payload lengths), relying on the feedback from the target model in the form of binary labels. The framework respects black-box conditions and relies on a predefined set of modifications, which ensures protocol-specific rules and semantic consistency, but also could limit the perturbation space that the agent can explore, thereby minimizing the effectiveness of the attack. Moreover, the authors do not provide a clear discussion on how these modifications preserve feature interdependencies. While the DRL-based attack effectively evades detection, the study does not fully explore how the DQN agent gathers information or maintains critical feature consistency.

TANTRA; a timing-based black-box adversarial attack has been proposed \cite{sharon2022tantra} to evade an anomaly-based NIDS (e.g., KitNET, Autoencoders, etc.) which rely on detecting deviations in traffic behavior. This attack is based on  reshaping the inter-packet delays of malicious traffic using a  LSTM model. Without direct access to the NIDS model’s internal workings, TANTRA learns temporal patterns from benign traffic and perturbs timing features without modifying packet content. It can also mimic legitimate timing patterns to avoid detection, sometimes completely evading detection systems. However, the authors do not explore how timing changes might impact other statistical features. 

To ensure realistic and valid traffic flows, Debicha et al. \cite{debicha2023adv} use a projection function to respect syntactic and semantic constraints during their proposed attack against flow-based NIDS. Their method leverages transferability by generating adversarial samples using surrogate models (e.g., MLP, RF, KNN) and perturbs important features like packet counts, transmission duration, and byte counts. Although it effectively evades detection, strategies for model querying and gathering target NIDS information are still unclear. Furthermore, relying on transferability has limitations, as its effectiveness depends on the similarity between the surrogate and target models, which may not be feasible in complex, real-world scenarios. To counteract the proposed attack, the authors propose an ensemble-based defense to detect adversarial samples. However, further evaluation against adaptive attacks was needed to validate its effectiveness in responding to evolving threats in dynamic real-world environment.


In \cite{lin2022idsgan}, the authors also used GAN to generate adversarial network traffic against different ML-based NIDS (i.e., SVM, RF, DT, etc.). It necessitates a generator to alter original malicious traffic into adversarial samples while maintaining the functional integrity of the attacks. Then, the discriminator learns from the real-time feedback of the target NIDS by repeatedly querying it. After this generator-discriminator training, IDSGAN exploits transferability of adversarial examples. It assumes that these attacks, which were fine-tuned on a local substitute model, will successfully evade detection by the target NIDS models. While the generated samples adhere to protocol rules, their reliance on repeated queries and transferability can limit real-world applicability. Moreover, the study focuses on functional features which lacks comprehensive validation of feature consistency and interdependencies.

Despite advancements and increased awareness of the importance of validity and feasibility in adversarial attack strategies \cite{apruzzese2022modeling, ennaji2024adversarial}, recent studies still ignore key factors like maintaining feature constraints and adhering to strict black-box conditions within network traffic. For example in \cite{roshan2024black}, the vulnerability of DNN-based NIDS to so-called black-box adversarial attacks has been evaluated based on transferability method. They use FGSM to create adversarial examples on the surrogate model, which is trained on the same dataset as the target model but with different hyperparameters. Their approach relies on assumptions of model similarity and data accessibility, which is incompatible with black-box settings. 

Another recent work by Zhang et al. \cite{zhang2024toward}, they used a linear autoencoder (LAE) as a surrogate model to improve cross-model transferability of adversarial attacks (e.g., R-FGSM, R-PGD, R-MIM, and Universal Adversarial Sample Generator (U-ASG)) to other AE-based detectors. While the study respects network protocol constraints, its focus on model similarity creates doubts about real-world feasibility. In addition, the paper's proposed methodology is limited by its lack of exploration of real-world black-box information gathering challenges.

Recently, He et al. \cite{he2024nids} proposed a novel decision boundary traversal algorithm namely NIDS-Vis, which visualizes and analyzes the decision boundaries of DNN-based NIDS to exploit its weaknesses. Based on experiments, the authors demonstrate that more complex models are easier to trick by adversarial inputs. To address this, they propose two methods: Feature Space Partitioning (FSP) to reduce clustering by dividing feature space, and a Distributional Loss Function (DLF) to align anomaly scores with a predefined distribution. The findings effectively visualizes decision boundaries to enhance resilience, but it lacks focus on real-world constraints and feature interdependencies. Moreover, the applicability of these methods in dynamic network scenarios is not thoroughly explored.

\begin{takeawaybox} Many proposed strategies ignore crucial constraints such as restricted model interactions and the absence of internal model knowledge, which are critical for practical and realistic black-box attacks. Despite claiming to operate in black-box settings, many studies including recent ones rely on assumptions about model access or data similarity, reducing their real-world feasibility. These methods often depend on repeated queries or transferability, making them easier to detect and misaligned with genuine attack scenarios. This exposes a significant gap in creating real and undetectable black-box adversarial techniques. Thus, no robust proposition for effective defenses adapted to industry needs \cite{kumar2020adversarial}. \end{takeawaybox}

\begin{table*}[ht]
\renewcommand{\tabcolsep}{2.55pt}
\renewcommand{\arraystretch}{1.2}
\centering
\caption{Related Works: Summary and Identified Limitations}
\resizebox{\textwidth}{!}{
\begin{tabular}{c|c|c|c|c|c|c|c|c|c}
\rowcolor{gray!40} \textbf{Reference} & \textbf{Year} & \multicolumn{2}{c|}{\textbf{Type of Attack}} & \textbf{Target Model} & \textbf{Dataset} & \multicolumn{4}{c}{\textbf{Limitations}} \\ 
\cline{3-4} \cline{7-10}

\rowcolor{gray!40} & & \textbf{White-Box} & \textbf{Black-Box} & & & \textbf{Real-World Feasibility} & \textbf{Feature Correlation} & \textbf{Black-Box Compliance} & \textbf{Information Gathering} \\ \hline

\rowcolor{gray!15} \cite{mohammadian2023gradient} & 2023 & \cmark & \xmark & MLP & 
\begin{tabular}[c]{@{}c@{}}CIC-IDS2017, \\ CIC-IDS2018, \\ CIC-DDoS2019 \end{tabular} 
& Low & \xmark & - & - \\ \hline

\cite{sheatsley2022adversarial} & 2022 & \cmark & \xmark & DNN & \begin{tabular}[c]{@{}l@{}}NSL-KDD,\\ UNSW-NB15 \end{tabular} & Medium & \cmark & - & - \\ \hline

\rowcolor{gray!15} \cite{anthi2021adversarial} & 2021  & \cmark & \xmark & DNN & Realistic ICS dataset & Low & \xmark & - & - \\ \hline

\cite{roshan2024untargeted} & 2024 & \cmark  & \xmark & DNN & CICIDS 2017 & Low & \xmark & - & - \\ \hline

\rowcolor{gray!15} \cite{roshan2024boosting} & 2024 & \cmark  & \xmark  & DNN  & CIC-DDoS-2019 & Low & \xmark & - & - \\ \hline

\cite{pawlicki2020defending} & 2020 & \centering \cmark & \xmark & DNN & \begin{tabular}[c]{@{}c@{}}NSL-KDD, \\ UNSW-NB15\end{tabular} & Low & \xmark & - & - \\ \hline

\rowcolor{gray!15} \cite{costa2024argan} &  2024 & \cmark & \xmark & kitNET  & Mirai attack  & Low & \xmark & - & - \\ \hline

\cite{zhang2022adversarial} & 2022 & \xmark & \cmark & \begin{tabular}[c]{@{}l@{}}CNN, DNN, \\ CNN+LSTM\end{tabular} & CSE-CIC-IDS2018 & Medium & \xmark & \xmark & \xmark \\ \hline

\rowcolor{gray!15} \cite{peng2019adversarial} & 2019 & \xmark & \cmark & DNN & \begin{tabular}[c]{@{}c@{}}KDDcup99, \\ CICIDS2017\\ \end{tabular} & High & \cmark & \cmark & \xmark \\ \hline

\cite{wu2019evading} & 2019 & \xmark & \cmark  & \begin{tabular}[c]{@{}l@{}}CNN, DT\end{tabular} & \begin{tabular}[c]{@{}c@{}}Malware Capture \\ Facility Project, \\ IOST 2010\end{tabular} & High & \cmark & \cmark & \xmark \\ \hline

\rowcolor{gray!15} \cite{sharon2022tantra} & 2022 & \xmark  & \cmark & \begin{tabular}[c]{@{}l@{}}KitNET, AE\end{tabular} & \begin{tabular}[c]{@{}l@{}}Kitsune,\\ CICIDS 2017\end{tabular} & High & \cmark & \cmark & \xmark \\ \hline

\cite{debicha2023adv} & 2023 & \xmark & \cmark & DNN &  \begin{tabular}[c]{@{}l@{}}CTU-13,\\ CICIDS 2017\end{tabular} & Medium & \cmark & \cmark & \xmark \\ \hline

\rowcolor{gray!15} \cite{zolbayar2022generating} & 2022  & \xmark & \cmark & DNN & \begin{tabular}[c]{@{}l@{}}NSL-KDD,\\ CICIDS2017\end{tabular} & Medium & \cmark & \xmark & \xmark \\ \hline

\cite{lin2022idsgan} & 2022 & \xmark & \cmark & \begin{tabular}[c]{@{}l@{}}SVM, RF, DT\end{tabular}  & NSL-KDD & Low & \xmark & \xmark & \xmark \\ \hline

\rowcolor{gray!15} \cite{roshan2024black} & 2024 & \xmark & \cmark & DNN & CICDDoS2019 & Low & \xmark & \xmark & \xmark \\ \hline

\cite{zhang2024toward} & 2024 & \xmark & \cmark & AE  & \begin{tabular}[c]{@{}l@{}}NSL-KDD,\\ CICIDS2017\end{tabular} & Medium & \xmark & \cmark & \xmark \\ \hline

\rowcolor{gray!15} \cite{he2024nids} & 2024 & \xmark & \cmark & DNN & UQ-IoT-IDS & Low & \xmark & \cmark & \xmark \\ \hline

\end{tabular}
}
\label{tab:gen_adv}
\end{table*}

\section{Proposed Black-Box Adversarial Vulnerability Assessment}
\begin{figure*}[h]
    \centering
    \includegraphics[width=1\textwidth]{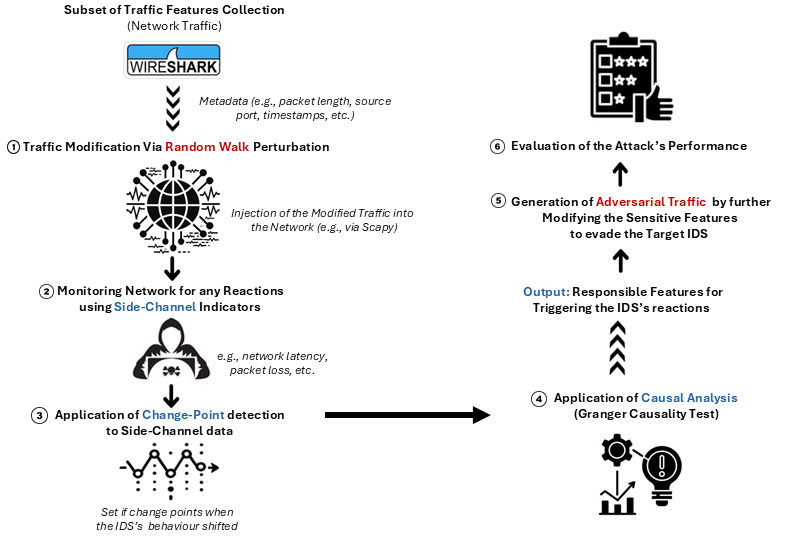} 
    \caption{Proposed Black-Box Adversarial Vulnerability Assessment}
    \label{fig:bb}
\end{figure*}

This section introduces our novel methodology for conducting an adaptive vulnerability assessment for black-box adversarial attacks. It also presents the target IDS model, describing its training process and the dataset used to ensure a realistic evaluation.
Our approach not only addresses several open research questions, but also directly helps industry by revealing practical adversarial threats, enabling the development of more effective defenses. Specifically, it clarifies how attackers could evade detection through silent probing, gather indirect information about the target IDS, and strategically select the most influential features for manipulation. To the best of our knowledge, our study is the first to provide a comprehensive framework for adaptive feature selection in a blind setting, using side-channel analysis, change-point detection, and causality analysis. The process consists of the following main steps, as illustrated in Figure \ref{fig:bb}.

\subsection{Target Model Training}
The target IDS model in this study was trained using a Random Forest (RF) classifier, a widely used machine learning model known for its resilience and interpretability within network security applications. The dataset employed for training was derived from the CSE-CIC-IDS2018 dataset, which offers a various range of real-world attack scenarios and normal network traffic patterns. A detailed description of the dataset is provided in Table \ref{tab:cse_cic_ids2018}.

\begin{table*}[ht]
\centering
\caption{Summary of the CSE-CIC-IDS2018 Dataset}
\label{tab:cse_cic_ids2018}
\renewcommand{\arraystretch}{1.2}
\rowcolors{2}{gray!15}{white}
\begin{tabular}{p{2.5cm}|p{15cm}}
\hline
\rowcolor{gray!40}
\textbf{Attribute}           & \textbf{Description} \\ \hline
\textbf{Traffic Type}        & Simulated network traffic covering multiple protocols, including HTTP, HTTPS, FTP, SSH, and email communications. \\ \hline
\textbf{Attack Types}    & Includes Brute Force (SSH, FTP), DoS (Slowloris, SlowHTTPTest, Hulk), DDoS (Botnet-based), Web Attacks (SQL Injection, XSS), Infiltration (Unauthorized Access), Botnet Traffic (Spam, Reconnaissance). \\ \hline
\textbf{Feature Categories}  & Consists of Basic Features (e.g., flow duration, total packets), Content Features (e.g., flags, flow statistics), Time Features (e.g., inter-arrival times), and Derived Features (e.g., bytes-per-second). \\ \hline
\textbf{Labeling}            & Each traffic flow is labeled as either benign or malicious, with specific attack categories available for granular analysis. \\ \hline
\textbf{Volume of Data}      & Over 16 million network flows collected across five days, capturing diverse attack scenarios. \\ \hline
\textbf{Total Features}  & 84 network-related attributes extracted for each flow, covering statistical, temporal, and protocol-based characteristics. \\ \hline
\end{tabular}
\end{table*}

\begin{itemize}
\item \textbf{Preprocessing and Feature Selection:} To maintain consistency during training, missing values were fixed by removing incomplete records. Feature normalization was applied by standardizing continuous attributes, ensuring uniformity across different scales. Finally, feature selection was performed by identifying and retaining the most significant network flow attributes, optimizing classification performance and minimizing computational cost.

\item \textbf{Model Training and Optimization:}
A Random Forest (RF) classifier was trained using hyperparameter tuning via grid search to optimize important parameters, including the number of estimators, maximum depth, and feature selection strategy. To avoid bias in learning, the training procedure ensured that the dataset composition was balanced, preserving an equal representation of malicious and benign traffic. Finally, performance validation was carried out to verify the IDS's capability to differentiate between legitimate and malicious network traffic using the standard evaluation metrics of accuracy, precision, recall, and F1-score.
\end{itemize}
These procedures were included in the trained target IDS's architecture, allowing it to identify a large range of malicious traffic patterns with high classification accuracy. The resilience of this model to our suggested silent probing adversarial attack is assessed and discussed in Section \ref{sec:discussion}

\subsection{Traffic Feature Collection}
Our methodology initiates with a traffic feature collection step based on network sniffing tools (e.g., Wireshark), to capture a subset of network metadata such as: 

\begin{itemize}
    \item \textbf{Packet length:} The size of each packet in bytes, which might indirectly provide certain traffic patterns or behaviours without exposing content.
    \item \textbf{Source port:} The originating port number of each packet, which provides information on the types of services or applications using the network.
    \item \textbf{Destination port:} The target port number that can be used to identify the type of communication, including common ports used by specific applications or protocols.
    \item \textbf{Timestamps:} The precise time each packet was delivered or received, providing information on timings and traffic patterns without revealing the contents of the packet.
    \item \textbf{Protocol type: } Without having access to payloads, the protocol being used (e.g., TCP/UDP) might provide an indirect indicator of communication characteristics.
\end{itemize}

By limiting the collected information to metadata only, our methodology respects black-box constraints, where attackers can only obtain data that is indirectly available and does not disclose the internal features or model parameters of the IDS. In contrast to conventional approaches that might assume some level of knowledge about the IDS’s dataset, feature space, or architecture, our method reflects the limitations an attacker faces in real-world scenarios, where access to privileged information is highly restricted.
This metadata serves as the foundation for our adaptive vulnerability assessment, allowing the adversary to observe general network behavior and identify potential indicators of IDS responses without requiring direct access.

\begin{algorithm}
\caption{Traffic feature collection}
\begin{algorithmic}[1]
\Procedure{TrafficFeatureCollection}{}
    \State Initialize \texttt{sniffer} (e.g., Wireshark)
    \State Create empty list \texttt{metadata\_list} to store extracted features
    \While{\texttt{sniffer} is active}
        \State Capture \texttt{packet} from network traffic
        \State Extract metadata fields from \texttt{packet} (e.g., length, ports, timestamp)
        \State Append extracted metadata to \texttt{metadata\_list}
    \EndWhile
    \State \Return \texttt{metadata\_list}
\EndProcedure
\end{algorithmic}
\end{algorithm}

\subsection{Traffic Modification via Random Walk Perturbations} 
In the second phase, random walk perturbations are applied to the collected traffic features, enabling silent probing that reduces detection risk. It involves three steps.
\begin{enumerate}
 \item \textbf{Perturbation initialization:} We define a small perturbation range, for each selected feature (e.g., packet length, inter-packet time intervals), which is limited to values that are common for normal traffic.
 \item \textbf{Random walk application:} To simulate natural shifts a slight, random value is added or subtracted for every feature within each packet. For example, in packet length, the packet size can be altered by adding/subtracting a few bytes, maintaining packet sizes within the expected range for the application.
 \item \textbf{Injection of the modified traffic:} The modified traffic is reintroduced into the network, enabling fine-grained control over packet-level attributes. Each packet modification is sufficiently minor to smoothly blend with the broader network traffic, making it indistinguishable from normal network activity.
\end{enumerate}
Unlike existing approaches that rely on repeated queries, our method subtly manipulate the metadata of network traffic in slight, randomized steps, allowing the attacker to assess the sensitivity of the target IDS to feature alterations without triggering alarms. The main purpose is to simulate natural, gradual variations in traffic patterns, similar to the minor deviations from standard patterns observed in legitimate network communications.

\begin{algorithm}
\caption{Traffic modification via random walk perturbations}
\begin{algorithmic}[1]
\Procedure{TrafficModification}{metadata\_list, modifiable\_features, base\_epsilon}
    \For{each feature $f$ in modifiable\_features}
        \State Set $\epsilon \leftarrow \text{base\_epsilon}$
        \For{$i = 1$ to num\_steps}
            \State $\text{perturbation} \sim \mathcal{N}(0, \epsilon \cdot \text{std}(f))$
            \State $f \leftarrow f + \text{perturbation}$
        \EndFor
    \EndFor
    \State Inject modified packets with perturbed features into the network
\EndProcedure
\end{algorithmic}
\end{algorithm}

\subsection{Monitoring Network Reactions Using Side-Channel Indicators} 
In situations when direct access to the system is not possible, side-channel indications allow an indirect method to infer the IDS's reactions to traffic disruptions, aligning with black-box constraints. To observe subtle variations in the IDS's behavior, we employ a set of side-channel metrics: response time, CPU usage, memory consumption, packet drop rate, and processing delay. These indicators are selected due to their relevance in capturing non-intrusive feedback from the IDS, enabling the attacker to assess system sensitivity without raising alarms.
\begin{itemize}
\item \textbf{Response time:} It shows how long it takes the IDS to process incoming packets. Response time-increasing perturbations may be a sign of increased processing demands, indicating that certain features are attracting more attention from the IDS.

\item \textbf{CPU usage:} By tracking CPU load fluctuations, we can infer the computational resources required by the IDS in response to specific traffic features. Increased CPU usage could indirectly indicate that specific packet properties are being examined. 

\item \textbf{Memory usage:} It provides information about how the IDS allocates its resources. By revealing which traffic modifications place a higher load on the IDS, variations in this statistic can indirectly highlight the difficulty of processing specific features.

\item \textbf{Packet drop rate:} Increased packet drop rates may indicate that the changed traffic is overloading or challenging the IDS. This response might bring attention to specific packet characteristics that the IDS finds difficult to manage in real time.

\item \textbf{Processing delay:} Slight but consistent delays in processing show that the IDS requires additional time to examine certain packets, which may be a sign that the manipulated features may be triggering more extensive analysis within the system.
\end{itemize}

\begin{algorithm}
\caption{Monitor network reactions using side-channel indicators}
\begin{algorithmic}[1]
\Procedure{MonitorReactions}{$\text{perturbedTraffic}$}
    \State Initialize metrics: $\text{latency}$, $\text{drop}$, $\text{CPU}$, $\text{memory}$, $\text{processing\_delay}$
    \For{each packet $p \in \text{perturbedTraffic}$}
        \State Inject $p$ into the network
        \State Measure side-channel metrics:
        \State \quad $\text{latency}(p)$, $\text{drop}(p)$
        \State \quad $\text{CPU}(p)$, $\text{memory}(p)$
        \State \quad $\text{processing\_delay}(p)$
        \State Store measurements in $\text{side\_channel\_data}$
    \EndFor
    \State \Return $\text{side\_channel\_data}$
\EndProcedure
\end{algorithmic}
\end{algorithm}

Existing research rarely explores the use of side-channel analysis in black-box adversarial attacks on NIDS. While they are a widely recognized method in cryptography for extracting sensitive information through indirect means (e.g., timing or power consumption), their application to NIDS has been limited. This is mainly because traditional NIDS research focus on direct analysis of network traffic, neglecting the potential of indirect system indicators. Our approach fills this gap by adapting side-channel strategies to the context of adversarial attacks against NIDS. Therefore, it enhances the understanding and realism of this evolving field and provides a foundation for developing defenses that address the risks of undetectable probing techniques.

\subsection{Adaptive Feature Identification with Constraint Adherence} 
Our methodology presents a novel approach to blindly identifying sensitive features that are most responsible for triggering responses in a NIDS. As discussed in Section \ref{sec:review}, most existing research on black-box adversarial attacks ignore the crucial aspect of feature selection. They often rely on direct manipulation of input data, frequently requiring knowledge of the model's internal structure or data. Moreover, most of them overlook the need for semantic and syntactic validity in modified features. Our approach addresses these limitations by combining change-point detection with causal analysis and carefully adhering to feature constraints at each phase of the process.

\textbf{Change-Point Detection: Definition and Application}
Change-point detection is a statistical technique employed to identify instances of structural changes in a time series \cite{niu2016multiple}. These shifts (i.e., change-points) indicate moments when the behavior of the observed system (in this case, the IDS) reacts significantly to external inputs. Detecting change-points isolates traffic modifications that correlate with substantial changes in the IDS's behavior, facilitating focused analysis of these features.

In this study, we apply the Binary Segmentation (Binseg) algorithm with different models, based on the characteristics of each side-channel indicator \cite{kovacs2023seeded}. For most indicators (e.g., CPU usage, memory consumption, and packet drop rate), we use the radial basis function (rbf) model, which accurately captures non-linear variations within the data and presents significant behavioral changes in reaction to traffic perturbations. For response time, we apply the l2 norm model, which is optimal for detecting minor linear shifts and gradual changes, indicative of potential packet inspection by the IDS.
These models were chosen to best reflect the distinct reaction across the different side-channel metrics, enabling a detailed analysis of IDS responses. Binary segmentation identifies 'change-points moments' where IDS behavior deviates from baseline due to traffic modifications. This allows us to pinpoint which traffic feature adjustments most strongly trigger IDS reactions. 

\begin{algorithm}
\caption{Adaptive feature identification}
\begin{algorithmic}[1]
\Procedure{AdaptiveFeatureIdentification}{$\text{side\_channel\_data}$, $\text{modifiable\_features}$}
    \State Initialize $\text{change\_points} \gets \{\}$ \Comment{To store change-points for each indicator}
    \State Initialize $\text{sensitive\_features} \gets \{\}$ \Comment{To store identified sensitive features}
    \For{each indicator $s$ in $\text{side\_channel\_data}$}
        \State Apply change-point detection on $s$ using Binary Segmentation (Binseg)
        \State $\text{breakpoints}_s \gets \text{Binseg}(s, \text{model} = \text{choose model})$
        \State Append $\text{breakpoints}_s$ to $\text{change\_points}$
    \EndFor
    \For{each segment $[t_i, t_{i+1}]$ between $\text{breakpoints}_s$}
        \State Extract segment $\text{data}_{[t_i, t_{i+1}]}$ for each $s$ and $\text{modifiable\_features}$
        \State \textbf{Perform Causal Analysis:}
        \State \quad Apply OLS regression: $y_s = \alpha + \sum_{f \in \text{modifiable\_features}} \beta_f f + \epsilon$
        \State \quad Apply Variance Inflation Factor (VIF) filter to ensure feature independence
        \State \quad Identify features with $p$-values $< \text{threshold}$ as significant
        \State Append significant features to $\text{sensitive\_features}$
    \EndFor
    \State \Return $\text{sensitive\_features}$
\EndProcedure
\end{algorithmic}
\end{algorithm}

\textbf{Causality Analysis: Definition and Application
}\\
Causality analysis \cite{runge2023causal} is a statistical method that shows how modifications to one variable (e.g., traffic features) impact another (i.e., IDS behavior). In the context of adversarial attacks, this allows attackers to identify and manipulate the most significant network traffic features, possibly evading IDS detection without fully knowing how it operates.

In our study, after identifying change-points in side-channel indicators, we apply causal analysis to verify which particular traffic features are most likely to be the cause of these reactions. This part of our approach involves a three-step process, ensuring statistical independence and causal validity in a black-box setting.
\begin{enumerate}
\item \textbf{Segment-based regression analysis:} For each segment identified between change-points, we use Ordinary Least Squares (OLS) regression \cite{zdaniuk2024ordinary} to estimate the coefficients of the linear regression model relating side-channel indicators to potential causal variables.
Without requiring direct IDS access, we can identify the traffic features that cause the most significant reactions from the IDS by examining each segment independently while remaining restricted by black-box requirements.
\item \textbf{Variance inflation factor filtering:} To preserve feature interdependencies, we calculate the Variance Inflation Factor (VIF) \cite{thompson2017extracting} for each feature in the regression model. Features with high VIF values, which may suggest multicollinearity and skew causal interpretations, are eliminated in this filtering stage. We ensure the causal analysis's robustness by retaining only statistically independent features, which effectively reflect the relationships within the traffic data.
\item \textbf{Features significance:} Sensitive features are those whose coefficients are statistically significant, as indicated by p-values below a predetermined threshold. A lower p-value indicates a stronger likelihood that the observed connection between a feature and the side-channel indicator reflects a real causal relationship rather than random chance. By prioritizing features with low p-values, the analysis focuses on those most likely to effectively affect the IDS behavior. This targeted approach optimizes subsequent analyses, enhancing the precision and relevance of our findings.
\begin{remarkbox} The choice of p-value threshold and its implications for feature selection are discussed in  detail in section \ref{sec:discussion}. \end{remarkbox}
\end{enumerate}

Having identified sensitive features through change-point detection and causal analysis, the groundwork is set for executing a targeted adversarial attack.

\subsection{Direct Adversarial Attack Generation}
After identifying sensitive features through causal analysis, our approach proceeds to a targeted adversarial attack phase. This requires injecting carefully controlled perturbations directly into these sensitive features, ensuring evasion of detection while preserving the realism of the network. The adversarial attack is carried out through the following steps.
\begin{itemize}
\item \textbf{Controlled perturbations:}  Perturbations are crafted within a narrow, controlled range to maintain protocol adherence and traffic plausibility, reducing the chance of triggering IDS alarms.
\item \textbf{Dynamic adjustment:} Based on feedback from initial probe outcomes, perturbation intensity can be adaptively adjusted, focusing on features that exhibit the highest sensitivity while maintaining expected traffic behavior.
\end{itemize}

\begin{algorithm}
\caption{Direct adversarial attack}
\begin{algorithmic}[1]
\Procedure{DirectAdversarialAttack}{$\text{sensitive\_features}$, $\text{traffic\_data}$, $\epsilon$, $\text{threshold}$}
    \State Initialize $\text{adversarial\_data} \gets \text{traffic\_data}$
    \For{each feature $f$ in $\text{sensitive\_features}$}
        \For{each sample $x$ in $\text{adversarial\_data}$}
            \State Compute perturbation: $\delta \gets \epsilon \times \text{std}(f)$
            \If{$\text{abs}(\delta) \leq \text{threshold}$} \Comment{Ensure perturbation within bounds}
                \State $x[f] \gets x[f] + \delta$
            \EndIf
        \EndFor
    \EndFor
    \State Inject modified adversarial samples into the network
    \State \Return $\text{adversarial\_data}$
\EndProcedure
\end{algorithmic}
\end{algorithm}

The following section presents a comprehensive evaluation of our approach's effectiveness, detailing its impact on IDS performance and assessing its undetectability in realistic network scenarios.

\section{Results and Discussion}
\label{sec:discussion}
To fill the gap between theory and practice, we evaluate IDS resilience under realistic black-box conditions, ensuring adaptability to industry settings for more resilient and effective defense strategies.
This section analyzes the effectiveness of the proposed approach and its real-world implications for IDS security. Compared to existing vulnerability assessments for adversarial attacks in the literature, our findings validate the feasibility of black-box adversarial attacks while revealing the limitations of current IDS defenses, which often fail to consider adversaries exploiting indirect observations rather than explicit model queries. 

The discussion covers: (i) the impact of silent probing on IDS, (ii) identification of sensitive features for adversarial manipulation, (iii) IDS performance degradation pre- and post-attack, and (iv) comparison with existing strategies.

\subsection{Evaluating the Effectiveness of Silent Probing}
The proposed silent probing strategy functions under strict black-box conditions, ensuring that the attacker does not require knowledge of the model's architecture, training data, or feature importance. In contrast to existing adversarial attack techniques that claim to work in black-box settings but rely on controlled environments, our approach successfully identifies sensitive features blindly through change-point detection and causal analysis without necessitating direct model queries. This distinction is crucial because it reflects real-world adversarial scenarios where attackers are unable to access the internals of the system.

As shown in algorithms 1 and 2, at the beginning of the attack process, we applied random walk perturbations to inject slight variations into the sniffed network traffic. For instance, features such as "Duration", "BytesPerSec", and "PktsPerSec" were perturbed using a slight epsilon value (0.01) over multiple steps. This facilitated an indirect analysis of the target IDS’s behaviour without triggering its detection mechanisms. 
Sniffing was performed beforehand to capture modifiable network traffic features, after which perturbed traffic was introduced to monitor side-channel indicators, such as "response\_time", "CPU\_usage", "memory\_usage", "packet\_drop", and "processing\_delay". After that, these indicators were carefully observed for any unusual alterations that would indicate sensitive system behaviours. This process provides necessary insights on the sensitivity of the target IDS subtle variations before we proceeded with more targeted adversarial activities.

\begin{figure*}[h]
    \centering
    \includegraphics[width=0.95\textwidth]{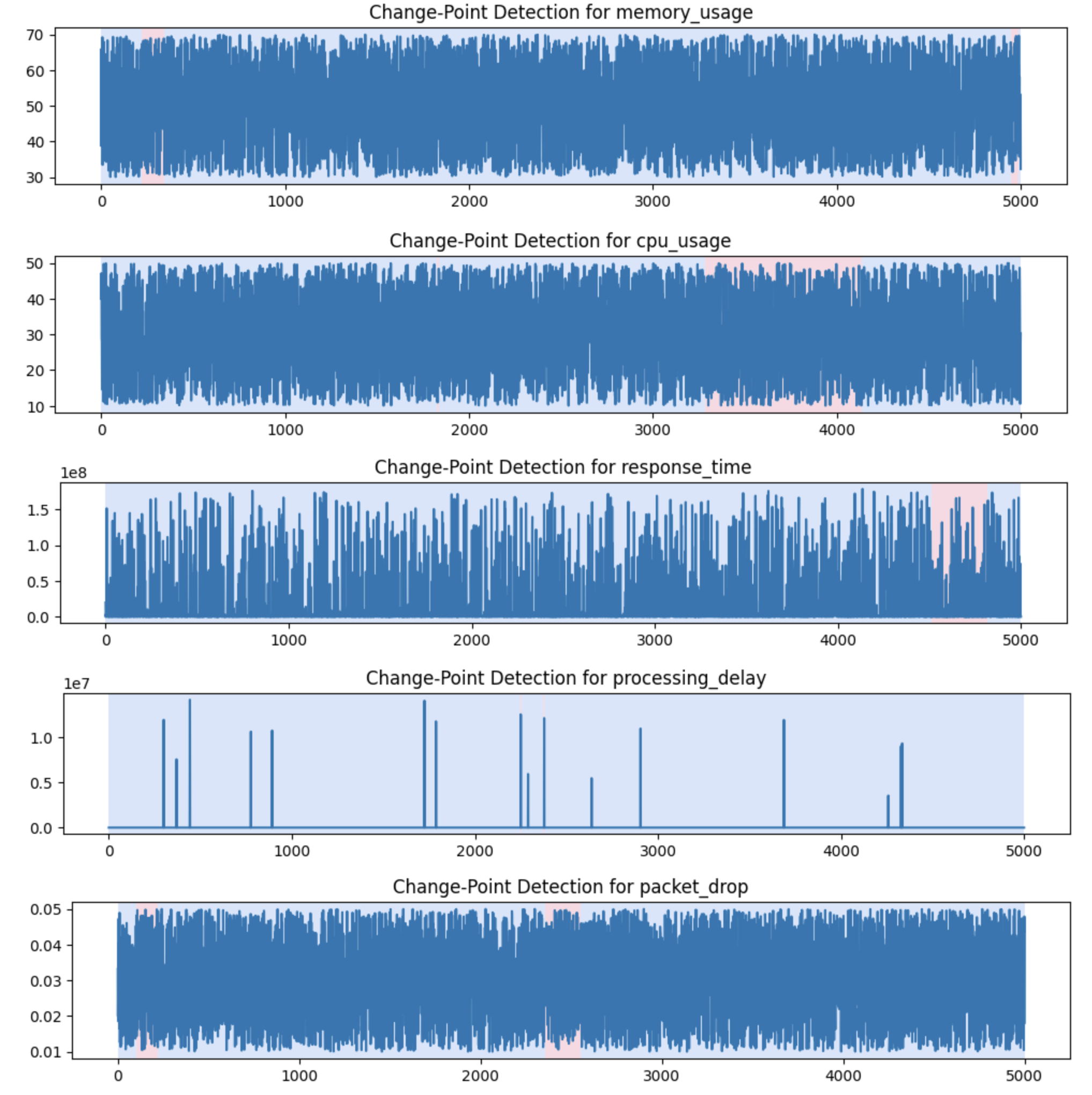} 
    \caption{Change-Point Detection in Side-Channel Indicators \textit{(memory usage, CPU usage, response time, processing delay, and packet drop)}.}
    \label{fig:cp}
\end{figure*}

The results obtained through change-point detection indicate significant shifts in side-channel metrics. These findings are visually supported by the change-point detection figure \ref{fig:cp}, which clearly identify intervals of significant variation in these metrics. 

\begin{itemize}
    \item \textbf{Response time:} exhibited abrupt variations at [3085, 3090, 4510, 4815, 5000], suggesting fluctuations in system processing behavior.
    \item \textbf{CPU usage:}  showed significant changes at [1825, 1835, 3285, 4135, 5000], reflecting potential spikes in resource allocation.
    \item \textbf{Memory usage:} had clear shifts at [220, 345, 4945, 4990, 5000], indicating possible fluctuations in workload handling.
    \item \textbf{Packet drop:} rates changed at [100, 220, 2355, 2550, 5000], suggesting congestion at these intervals.
    \item \textbf{Processing delay:} exhibited major transitions at [2250, 2255, 2375, 2380, 5000], highlighting moments of system slowdowns.
\end{itemize}

However, while change-point detection highlights where abrupt changes occur, it does not provide insights into what causes these changes. To address this, we leveraged causal analysis to determine which network traffic features were responsible for these variations.

\subsection{Identification of Sensitive Features for Direct Adversarial Manipulation}

\begin{figure*}[h]
    \centering
    \includegraphics[width=0.97\textwidth]{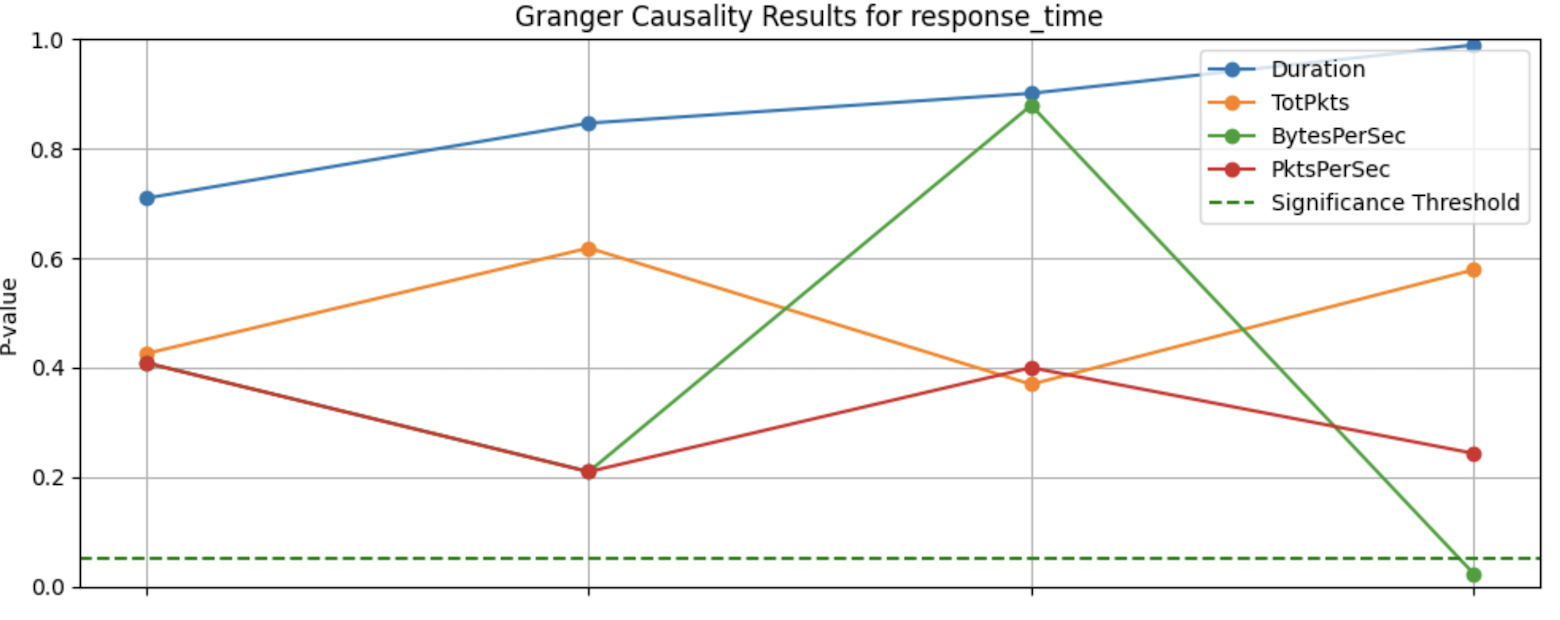} 
    \caption{Granger Causality Results for Response Time}
    \label{fig:1}
\end{figure*}

\begin{figure*}[h]
    \centering
    \includegraphics[width=0.97\textwidth]{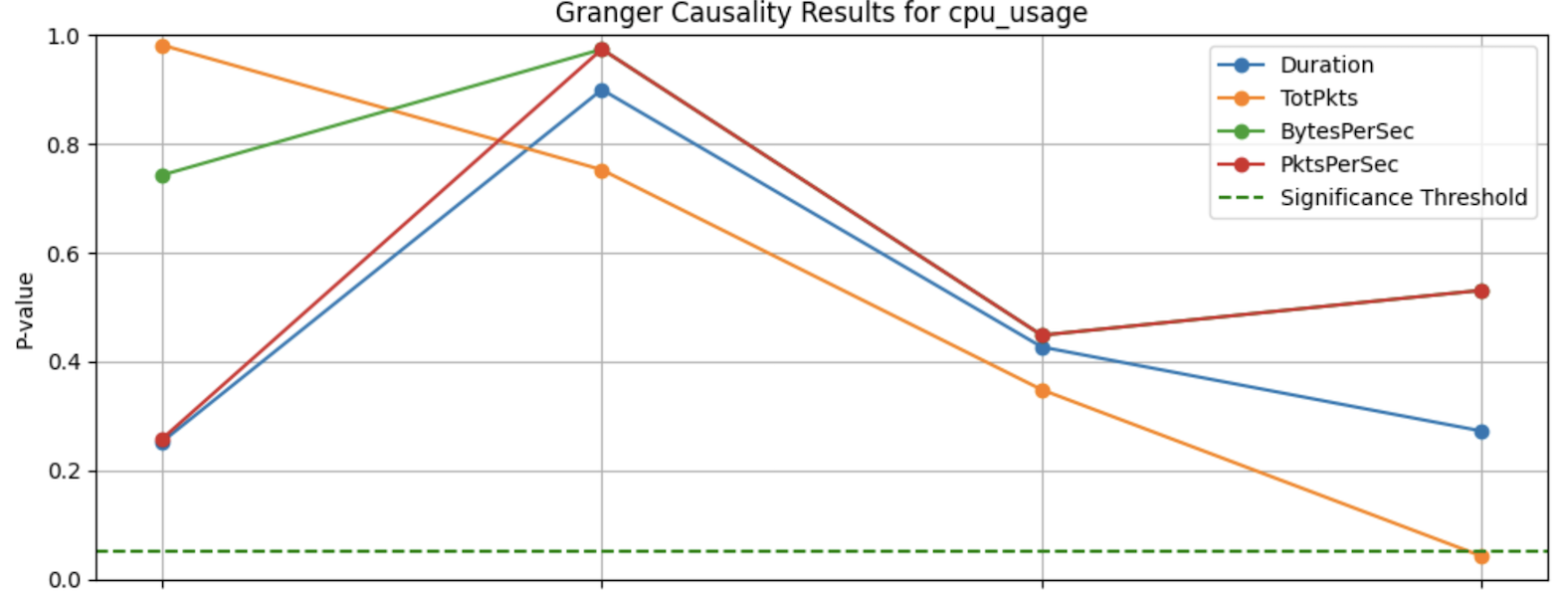} 
    \caption{Granger Causality Results for CPU Usage}
    \label{fig:4}
\end{figure*}

\begin{figure*}[h]
    \centering
    \includegraphics[width=0.97\textwidth]{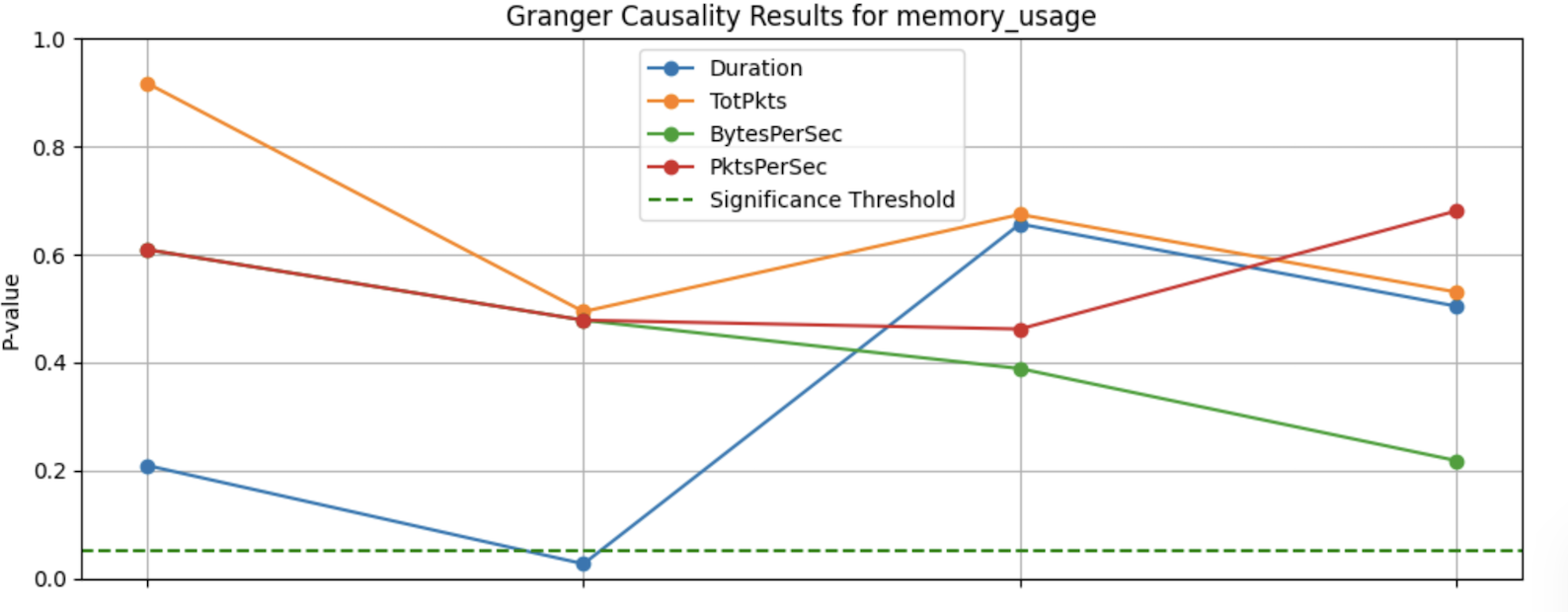} 
    \caption{Granger Causality Results for Memory Usage}
    \label{fig:2}
\end{figure*}

\begin{figure*}[h]
    \centering
    \includegraphics[width=0.97\textwidth]{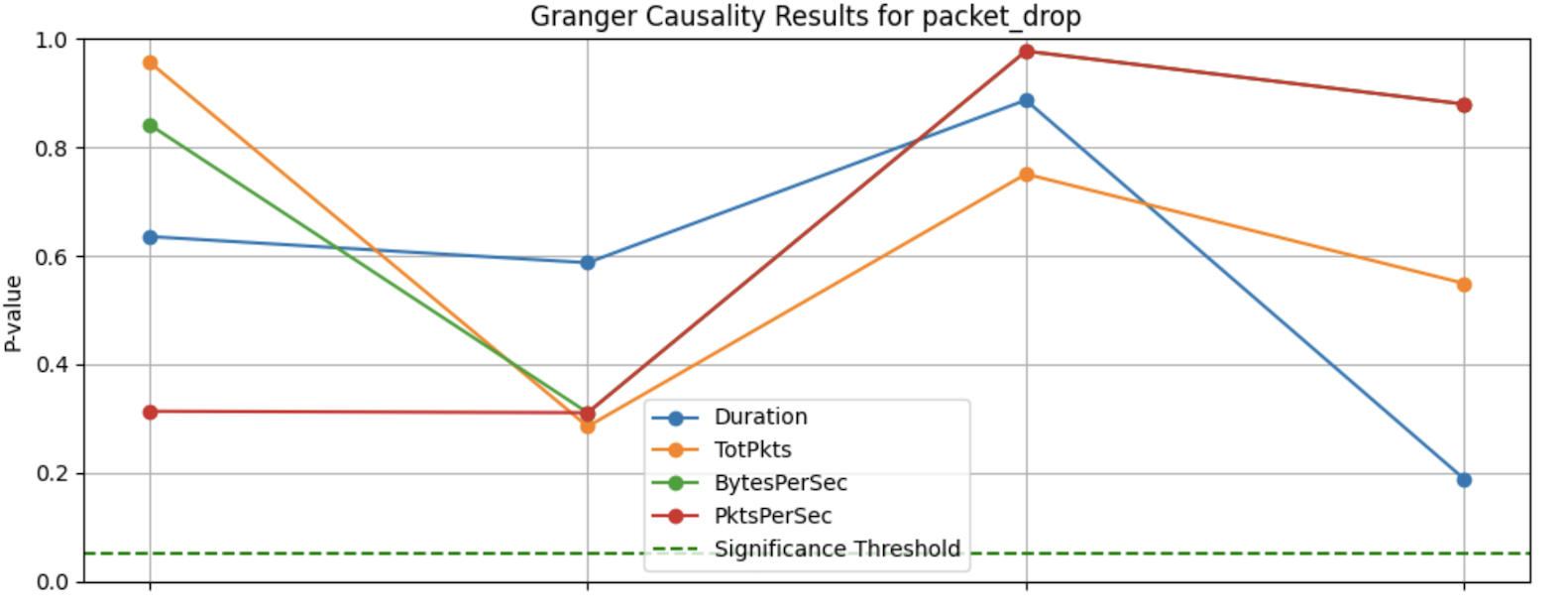} 
    \caption{Granger Causality Results for Packet Drop}
    \label{fig:3}
\end{figure*}

\begin{figure*}[h]
    \centering
    \includegraphics[width=0.97\textwidth]{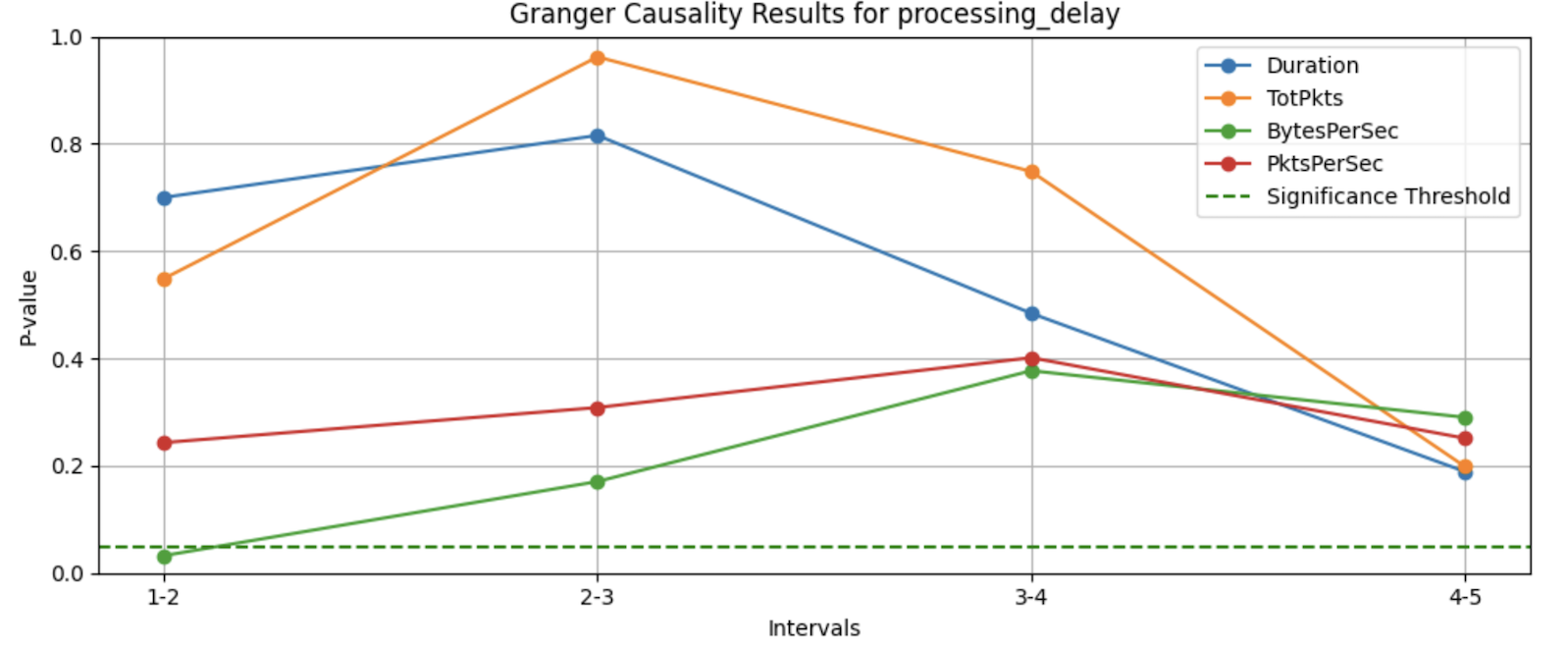} 
    \caption{Granger Causality Results for Processing Delay}
    \label{fig:5}
\end{figure*}

To further understand the root cause of the observed side-channel shifts, we used Granger causality tests for causal analysis. As presented, in figures \ref{fig:1}, \ref{fig:2}, \ref{fig:3}, \ref{fig:4}, and \ref{fig:5}, main traffic features that influenced system behavior at various intervals.

\textbf{Response time:} As illustrated in Figure \ref{fig:1}, According to the outcomes of the causal analysis, "BytesPerSec" significantly impacted the response time, especially during the later intervals. At intervals 4-5, the p-values for "BytesPerSec" fell notably below the significance threshold (0.05), demonstrating that shifts in traffic volume had a direct impact on latency. Similarly, "PktsPerSec" presents a strong decline in p-values around interval 3-4, showing a clear causal link.

\textbf{CPU Usage:} In Figure \ref{fig:4}, BytesPerSec and PktsPerSec have been shown to be important factors that affect processor load in the causal analysis of CPU usage.   At intervals 2-3, the p-value for "PktsPerSec" was particularly high (p = 0.24), which was consistent with the breakpoints identified by change-point detection. At intervals 3–4, "BytesPerSec" reached a significant increase, confirming its impact on computational load.

\textbf{Memory usage:} Figure \ref{fig:2} depicts that "Duration" and "TotPkts" have been identified to be the main contributors to the observed deviations in memory usage. "Duration"'s impact on memory fluctuations was confirmed by the constantly low p-value, which dropped immediately below the significance threshold in intervals 2-3. Furthermore, "TotPkts" showed periodic influence, especially in intervals 3–4, which is consistent with the theory that memory strain is caused by persistently high packet volumes.

\textbf{Packet drop:} Figure \ref{fig:3} shows that "PktsPerSec" and "TotPkts" are the dominant influencing factors in the packet drop Granger causality test. With p-values falling below 0.05 in intervals 2-3 and 4-5, PktsPerSec demonstrated significant causation and a definite influence on network congestion. Similarly, "TotPkts" presented a peak in influence during interval 3-4, synchronizing with sudden changes in dropped packets observed in the change-point analysis.

\textbf{Processing delay:} The analysis shown in \ref{fig:5} that "Duration" and "BytesPerSec" were significantly correlated with processing delay. The p-value for "BytesPerSec" reached its lowest value in interval 2-3, indicating that high traffic throughput introduced noticeable processing slowdowns. Intervals 3–4 and 4-5 revealed increases in "Duration", confirming its impact on traffic delays during extended sessions.

By analyzing side-channel variations using change-point detection and causal analysis for key traffic features, the most sensitive features triggering IDS detection changes that we identified are: 
\textbf{"Duration", "BytesPerSec", "PktsPerSec", and "TotPkts"}.
These features demonstrate strong correlations with \textbf{response time, CPU usage, memory usage, packet drops, and processing delays}, making them the optimal target for efficient adversarial perturbations.
\begin{tcolorbox}[
    colback=gray!10,   
    colframe=gray!40,  
    title=\textbf{TAKEAWAY},
    fonttitle=\bfseries
]
The proposed vulnerability assessment framework remains realistic and respect black-box conditions, ensuring that no prior knowledge of the IDS internals is needed. This reflects its applicability in real-world scenarios. In the next step, we discuss the findings after generating direct adversarial perturbations on the selected features, systematically degrading IDS performance while remaining undetected.
\end{tcolorbox}

\subsection{IDS Performance Pre- and Post-Attack}
Unlike existing adversarial attacks that rely on query-based methods or surrogate models, which often do not present realistic scenarios \cite{}, our attack operates in a fully black-box environment, exploiting side-channel indicators to guide direct feature perturbations.

To prevent detection, we used progressive feature perturbations, ensuring they maintained within statistical noise thresholds.
Instead of direct gradient-based techniques \cite{}(which necessitate model access), we adopted a stealthy perturbation strategy, gradually changing feature values over a number of steps ($num\_steps=75$). These manipulations were not random, they were guided by system responses, particularly the observed variations in side-channel behaviors (processing delay, response time, etc.). To boost attack effectiveness while avoiding detection by anomaly-based defenses, we progressively increased the perturbation intensity ($\epsilon = 0.15$).

\begin{figure*}[h]
    \centering
    \includegraphics[width=0.95\textwidth]{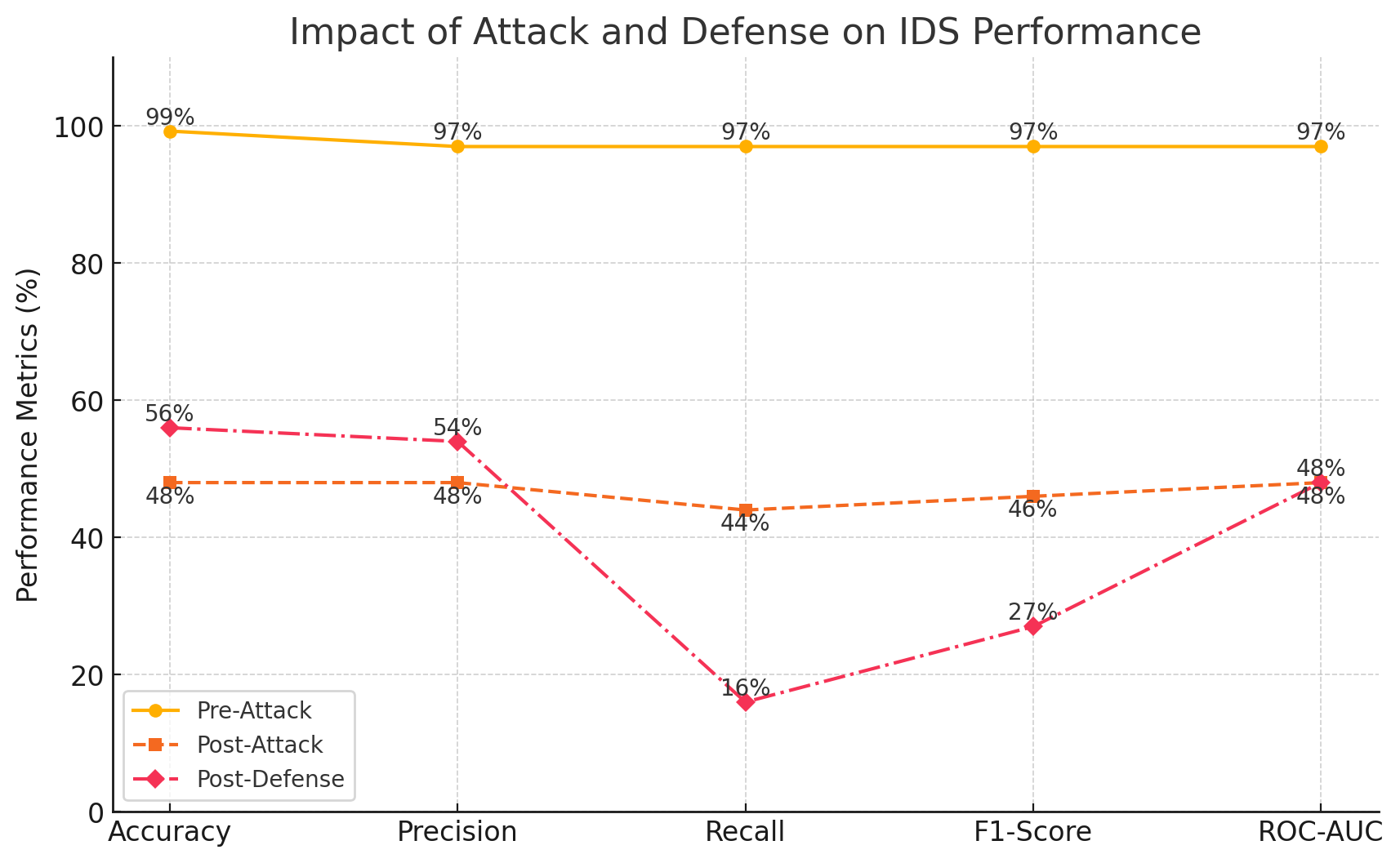} 
    \caption{Impact of Adversarial Attack on the Orginal Performance of the IDS \textit{(Pre- and Post-Attack)} and Post-Defense IDS Performance Metrics}
    \label{fig:defense}
\end{figure*}

As shown in Table \ref{tab:ids_performance} and Figure \ref{fig:defense}, the \textbf{Accuracy} of the IDS dropped from 99.25\% to 48.00\%, effectively reducing its classification capability to almost random behavior. This severe decline suggests that the IDS is no longer reliable to distinguish between benign and attack traffic, which makes its predictions highly unaccurate. 
\begin{table}[h]
    \centering
    \renewcommand{\arraystretch}{1.3} 
    \setlength{\tabcolsep}{10pt} 
    \rowcolors{2}{gray!15}{white} 
    \caption{Impact of Adversarial Attack on IDS Performance.}
    \begin{tabular}{c|c|c|c}
        \hline
        \rowcolor{gray!40} \textbf{Metric} & \textbf{Pre-Attack} & \textbf{Post-Attack} & \textbf{Change (\%)} \\
        \hline
        \textbf{Accuracy} & 99.25\% & 48\% & \textbf{-51.25\%} \\
        \hline
        \textbf{Precision} & 97\% & 48\% & \textbf{-50.52\%} \\
        \hline
        \textbf{Recall} & 97\% & 44\% & \textbf{-54.64\%} \\
        \hline
        \textbf{F1-score} & 97\% & 46\% & \textbf{-52.58\%} \\
        \hline
    \end{tabular}
    \label{tab:ids_performance}
\end{table}
This drop in performance is particularly concerning as it shows that even a well-trained IDS, which initially provided efficient classification results, is susceptible to adversarial perturbations when crafted under realistic black-box constraints. 

\textbf{Precision} describes the ratio of correctly detected attack  instances among all samples classified as attacks. The significant 50.52\% drop indicates that the IDS wrongly detects an important number of benign traffic events as attacks.

Regarding the outcomes of \textbf{Recall}, which measures the proportion of actual attacks correctly detected by the IDS,  it has dropped from 97\% to 44\% (54.64\% drop). This indicates that the IDS is missing more than half of the actual attacks within the network traffic. 

The trade-off between accurately identifying attacks \textbf{(Recall)} and reducing false alarms \textbf{(Precision)} is balanced by the \textbf{F1-score}, which is the harmonic mean of accuracy and recall. The IDS's overall failure in both aspects is highlighted by a drop from 97\% to 46\%.

\begin{tcolorbox}[
    colback=gray!10,   
    colframe=gray!40,  
    title=\textbf{TAKEAWAY},
    fonttitle=\bfseries
]
From a cybersecurity viewpoint, This raises the false positives as well as false negatives, which in real-world deployments may have serious implications:
\begin{itemize}
\item \textbf{Increased false positives:} Excessive alerts fatigue and ineffective incident response can result from network administrators receiving many alarms.
\item \textbf{Decreased trust in IDS:} If the IDS frequently misclassifies benign traffic as an attack, operators might ignore real attack warnings, affecting network security.
\item \textbf{Evading detection:} With a Recall value of 44\%, over 55\% of malicious traffic is not detected, which makes it simpler for adversaries to get into the system. 
\end{itemize}
\end{tcolorbox}

To validate the undetectability of our silent probing approach and evaluate whether it accurately reflects realistic adversarial scenarios, an Isolation Forest (IF) anomaly detection technique has been employed. It is an unsupervised anomaly detection algorithm \cite{xu2023deep} that isolates anomalies instead of profiling normal patterns. By randomly dividing the feature space, it creates several decision trees; because anomalies (outliers) are unusual, it is expected that they will be isolated in reduced splits. The lower the number of partitions needed to isolate a sample, the more likely it is that it will be considered abnormal. This makes IF an effective method for detecting statistically abnormal traffic patterns.

For our experiments, we trained the IF on both benign and adversarial traffic using the side-channel indicators (response time, CPU usage, memory usage, packet drop, and processing delay) along with the extracted sensitive features from causal analysis ("Duration", "BytesPerSec", and "PktsPerSec"). If our silent probing attack was truly undetectable, this defense strategy should fail to differentiate between benign traffic and adversarially perturbed one.

\begin{table}[h]
    \centering
    \renewcommand{\arraystretch}{1.3}
    \rowcolors{2}{gray!15}{white}
    \caption{Detection Performance of the Isolation Forest Anomaly Detector to the Proposed Silent Probing Approach.}
    \resizebox{0.3\textwidth}{!}{  
    \begin{tabular}{l c}
        \rowcolor{gray!40}
        \textbf{Metric} & \textbf{Post-Defense Value} \\
        \hline
        Detection Accuracy & 56.00\% \\
        Precision & 54.00\% \\
        Recall & 16.00\% \\
        F1-Score & 27.00\% \\
        ROC-AUC & 48.00\% \\
    \end{tabular}
    }
    \label{tab:undetectability_results}
\end{table}

The low detection performance, across all key metrics, is presented in Table \ref{tab:undetectability_results} and Figure \ref{fig:defense}. The accuracy only increased from 48\% to 56\%, indicating that the IF method failed to correctly classify benign and adversarial traffic. 

Moreover, the precision value of 54\% explains that almost half of the traffic events flagged as malicious by the anomaly detector were actually benign traffic.

Additionally, the recall indicates that only 16\% of adversarial traffic was correctly identified as an attack, while 84\% of adversarial examples evaded detection.

The imbalance presented by F1-Score (27\%) confirms that despite the detectability of some attacks, the overall effectiveness of the defense is poor due to its inability to balance detection accuracy.

Finally, the ROC-AUC score of 48\% implies that the classifier's random classification at 50\%, which differentiates abnormal traffic from normal one, is just slightly better than flipping a coin. This demonstrates that, in a real-world scenario, our attack successfully remains undetected.

\begin{tcolorbox}[colframe=gray!30, colback=gray!10, title=TAKEAWAY, fonttitle=\bfseries] Our experimental results validate that our silent probing strategy remains strongly undetectable, reflecting its high feasibility in realistic environments where attackers cannot rely on explicit model queries.
\end{tcolorbox}

\begin{table*}[h]
    \centering
    \renewcommand{\arraystretch}{1.3}
    \rowcolors{2}{gray!15}{white}
    \caption{Comparison of Our Approach with Existing Black-Box Adversarial Attack Strategies}
    \begin{tabular}{p{4.5cm} p{5.5cm} p{5.5cm}}
        \rowcolor{gray!40}
        \textbf{Comparison Criteria} & \textbf{Existing Black-Box Attacks} & \textbf{Our Approach} \\
        \hline
        \textbf{Black-Box Assumptions} & Often assume surrogate models or partial knowledge of feature space. & Fully black-box, no assumptions about feature space or model architecture. \\
        \textbf{Query Interaction} & Require repeated queries to extract decision boundaries. & \textbf{Minimal interaction}, relies on passive side-channel observations. \\
        \textbf{Feature Selection} & Heuristic-based or predefined, often requiring manual tuning. & \textbf{Adaptive}, leveraging change-point detection and causal analysis. \\
        \textbf{Empirical Validation} & Evaluated in controlled settings with constrained attack scenarios. & \textbf{Systematically validated}, ensuring attack feasibility under real-world constraints. \\
        \textbf{Computational Cost} & High overhead due to iterative query-based attack strategies. & \textbf{Lightweight}, does not require high query complexity. \\
        \textbf{Real-World Applicability} & Limited; many methods assume controlled data distributions or synthetic environments. & \textbf{High}; attack framework is designed for practical deployment scenarios. \\
        \textbf{Detectability Risk} & Higher risk due to reliance on direct model interaction, making attacks more noticeable. & \textbf{Low risk}, designed for silent probing and near-undetectability by anomaly detection systems. \\
    \end{tabular}
    \label{tab:comparison_results}
\end{table*}

\subsection{Comparison to Existing Approaches}

As mentioned in Table \ref{tab:vulnerability_methods}, the vulnerability assessment methods for conducting adversarial attacks against IDS can generally be categorized into query-based, transferability-based, decision boundary-based, and randomized attacks. This section compares these four categories to existing our proposed approach, evaluating query dependency, feature selection, efficiency, and real-world feasibility.  We highlight previous limitations and show that our method provides a more practical, stealthy, and adaptive vulnerability assessment.

Query-based methods, such as the ZOO attack \cite{chen2017zoo} and OPT attack \cite{cheng2019sign}, require frequent interaction with the target model to approximate its gradients.  Although these methods have shown to be effective, their high query costs and increased detection risks from IDS logging mechanisms make them completely impractical in real-world black-box scenarios.

Transferability-based attacks, such as FGSM (Fast Gradient Sign Method) \cite{goodfellow2014explaining} and C\&W \cite{carlini2017towards}, claim that the target IDS can be successfully tricked by an adversarial example generated on a surrogate model. However, because feature distributions and model structures vary throughout networks, this assumption frequently fails in real-world situations.   Transfer-based attacks have been shown to be ineffective against sophisticated IDS models, especially those that use adaptive training or ensemble learning \cite{apruzzese2022modeling}.

Decision boundary-based attacks, including Boundary Attack \cite{brendel2017decision} and HopSkipJump \cite{chen2020hopskipjumpattack}, iteratively refine adversarial examples to exceed classification thresholds. These techniques work effectively against classifiers with well-defined boundaries, but they are time-consuming and computationally expensive, which limits their applicability in real-time adversarial situations.

Randomized \& gradient-free attacks, such as GenAttack \cite{alzantot2019genattack}, rely on evolutionary algorithms \cite{bartz2014evolutionary} to create adversarial manipulations without explicit gradient access. Despite enhancing undetectability, these strategies require extensive sampling and statistical modeling, which may not align with real-time network conditions.

In contrast, our silent probing approach eliminates these limitations by removing the need for direct model interaction. Our approach uses passive side-channel indicators such processing latency, CPU usage, memory usage and packet drop rates to infer model behavior rather than querying the IDS.  This method ensures ideal feasibility in real-world deployments while adhering to strict black-box constraints. Unlike previous methodologies and the ones discussed in Section \ref{sec:review} (\cite{debicha2023adv, zhang2022adversarial, peng2019adversarial}), which often assume access to model outputs, decision thresholds, or training data distributions, our attack remains stealthy and undetectable, significantly validating its practical applicability.

Table \ref{tab:comparison_results} summarizes the comparison between existing black-box adversarial attacks and our approach, reinforcing the novelty and real-world effectiveness of our study. These findings lay the groundwork for future studies to developing adaptive adversarial defenses, ensuring that IDS models can proactively counteract evolving attack strategies while remaining resilient in real-world applications.

\section{Conclusion and Future Scope}\label{sec:conclusions}

This study presents a novel silent probing-based adversarial attack that reveal crucial vulnerabilities in Intrusion Detection Systems (IDS) under black-box conditions. Unlike existing adversarial techniques that assume impractical knowledge of model internals or lack transparency in their vulnerability assessment methods, our approach adhere to realistic attacker conditions. Instead of relying on direct queries, we rely on side-channel information, change point detection, and causal analysis in order to improve both feasibility and understanding in this field, paving the path for reliable security measures and stronger defenses. Through change-point detection and causal analysis, we showed that it is possible to identify the most sensitive network traffic features and strategically apply adversarial manipulations to decrease the performance of the target IDS performance. Our findings show a significant drop in IDS accuracy from 99.25\% to 48\%, effectively canceling its detection capability. Moreover, we used an Isolation Forest-based anomaly detector that validate the undetectability of the proposed attack. The post-defense accuracy remained low at 56\%, with an F1-score of just 27\%, exposing how adversarial probing attacks cannot be mitigated by current anomaly detection methods. These results highlight how urgently IDS deployments require stronger, more adaptive security measures.

Importantly, the overall objective of this work is to fill the gap between research on theoretical adversarial attacks and real cybersecurity issues. We facilitate the advancement of industry-ready defenses that proactively minimize adversary risks without affecting system efficiency by revealing weaknesses in IDS frameworks.





\end{document}